\documentclass[a4paper]{article}
\usepackage{graphicx}
\usepackage{amssymb}
\usepackage{amsfonts}

\title{Affine Quantization on the Half Line}

\author{Laure Gouba \\
Abdus Salam International Centre for Theoretical Physics,                       ICTP \\ Strada Costiera, 11, I-34151 Trieste Italy\\
Email: lgouba@ictp.it}

\begin{document}

\maketitle

\begin{abstract}
The similarity between classical and quantum physics is large enough to make an investigation of quantization methods a worthwhile endea\-vour. As history has shown, Dirac's canonical quantization method works reasonably well in the case of conventional quantum mechanics over $\mathbb{R}^n$ but it may fail in non-trivial phase spaces and also suffer from ordering problems. Affine quantization is an alternative method, similar to the canonical quantization, that may offer a positive result in situations for which canonical quantization fails. In this paper we revisit the affine quantization method on the half-line. We formulate and solve some simple models, the free particle and the harmonic oscillator. 
\end{abstract}

\section{Introduction}

Although non-relativistic quantum mechanics stands as a well- established theory and a well experimentally test theory,  the question of how to pass from classical to quantum theory  and a better understanding of the relation between classical and quantum mechanics is still of particular interest. Indeed, 
\begin{itemize}
\item ongoing attempts to quantize general relativity where a definitive answer to the question of the correct quantum theory of gravitation is still missing,
\item a quantization method that is able to take the nonlinear structure into account right from the outset is a useful tool to construct and study possible candidates for a theory of gravity,
\item a better knowledge of quantization in a situation when physical systems satisfy constraints or boundary conditions is needed.
\end{itemize}

In physics, quantization is generally understood as a  correspondence between a classical and a quantum theory. The question is how can we construct a quantum theory if a classical system is given ?
If we consider the quantum theory to be a more fundamental theory and classical mechanics to be only approximatively correct, the very concept of quantization seems pointless or appears to be ill-founded since it attempts to construct a {\it correct}  theory from a theory which is only approximatively {\it correct}. For instance there is the phenomenon of classical anomaly in the sense that if 
the quantization of some models gives the necessary quantum 
system, it may turn out that not all the quantum symmetry properties of the system are automatically reflected at the
classical level \cite{mikha1}. 
There are quantum systems for which no classical counterpart exists: for example He-II superfluidity and some broad class of such systems can be found in the literature \cite{mikha2}.

Quantum mechanics, like any other physical theory, classical mechanics, electrodynamics, relativity, thermodynamics, cannot be derived. The laws of quantum mechanics, expressed in mathematical form, are the results of deep physical intuition, as indeed, are all other physical theories. Their validity can only be checked experimentally. From this point of view, quantization is not a method for deriving quantum mechanics, rather is a way to understand the deeper physical reality which underlies the structure of both the classical and quantum mechanics and which unifies the two from geometrical perspectives.

It is conceptually very difficult to describe a quantum theory from scratch, without the help of a reference classical theory. The similarity between classical and quantum physics is large enough to make quantization a worthwhile approach. There is a certain mathematical richness in the various theories of quantization where the method does make sense.
Quantization in its modern sense is therefore often understood as the construction of a quantum theory with the help of a classical reference, not necessarily as a strict mapping. 
Quantization is studied not only for the sake of novel predictions: it is equally rewarding to reproduce existing results in a more illuminating manner.

Originally P. A. M. Dirac introduced the canonical quantization in his $1926$ doctoral thesis, {\it The method of classical analogy for quantization} \cite{pamd}. The ca\-no\-ni\-cal quantization or correspondence principle is an attempt to take a classical theory described by the phase space variables, let's say $p$ and $q$, and a Hamiltonian $H(q,p)$ to define or construct its corresponding quantum theory. The following simple technique for quantizing a classical system is used. Let $q^i, \: p_i, i= 1,2\ldots n$, be the canonical positions and momenta for a classical system with
 $n$ degrees of freedom. 
Their quantized counterparts 
$\hat q^i, \hat p_i,\: i= 1,2,\ldots n$,  are to be realized as operators on the Hilbert space $\mathcal{H} = L^2(\mathbb{R}^n, dx)$ by the prescription 
\begin{equation}
(\hat q^i\psi)(x) = x^i\psi(x);\quad 
(\hat p_i\psi)(x) = -i\hbar\frac{\partial}{\partial x^i}\psi(x);\quad i= 1,2,\ldots n,\quad x\in \mathbb{R}^n.
\end{equation}

This method is known as canonical quantization and is the basic method of quantization of a classical mechanics model 
\cite{dir, wey, vneu}. More general quantities, such as the Hamiltonians, become operators according to the rule 
\begin{equation}
H(p, q) \rightarrow  \hat H(\hat p , \hat q), 
\end{equation}
an expression that may have ordering ambiguities \cite{born, agar}.
In which canonical coordinates system does such a quantization method works ?
\begin{enumerate}
\item According to Dirac, replacing classical canonical coordinates by corresponding operators is found in practice to be successful only when applied with the dynamical coordinates and momenta referring to a cartesian system of axes and not to more general curvilinear coordinates.
\item Cartesian coordinates can only exist on a flat space.
\item The canonical quantization seems to depend on the choice of coordinates.
\item Beyond the ordering problem, one should keep in mind that 
$[\hat q,\; \hat p] = i\hbar\,I_d$ holds true with self adjoint operators $\hat q,\;\hat p$, only if both have continuous 
spectrum ($-\infty, +\infty$), and there is uniqueness of the solution, up to unitary equivalence (von Neumann).
\end{enumerate}
There are two attitudes that may be taken towards this apparent dependence of the method of the canonical quantization on the choice of coordinates. 
The first view would be to acknowledge the cartesian character that is seemingly part of the method. The second view would be to regard it as provisional and seek to find a quantization formulation that eliminates this apparently unphysical feature of the current approaches.
 
The aim of eliminating the dependence on cartesian coordinates in the standard approaches is no doubt one of the motivations for several methods such as the geometric quantization \cite{geo1, geo2, geo3, geo4}, the path integral quantization \cite{fey}, the deformation quantization \cite{geo2, def2, def3, def4, deff, def5}, the Klauder-Berezin-Toeplitz quantization \cite{kbt1, kbt2, ber, kbt3}. 
There is no general theory of quantization presently available which is applicable in all cases, and indeed, often the techniques used to quantize has to be tailored to the problem in question.

As history has shown, Dirac's canonical quantization method works reasonably well in the case of conventional quantum mechanics over $\mathbb{R}^n $ due to the following reasons :
\begin{itemize}
\item the underlying configuration space $\mathbb{R}^n$ is so well behaved;
\item when we try to quantize classical systems with phase spaces other than the cotangent bundle $T^\star\mathbb{R}^n$, the situation changes drastically;
\item already in classical mechanics, the phase spaces different from $T^\star\mathbb{R}^n$ require a more elaborated mathematical formalism;
\item global and topological aspects play a much bigger role in quantum theory than in classical physics.
\end{itemize}

Although quite successful in applications, the canonical quantization method has some severe shortcomings from a theoretical point of view.
A number of questions arise in connection with the scheme of canonical quantization. 
\begin{enumerate}
\item Let Q be the position space manifold of the classical system and q any point in it. Geometrically, the phase space of the system is the cotangent bundle $\Gamma = T^\star Q$. If Q is linear, means $ Q \sim \mathbb{R}^n$, then the replacement $q^i \rightarrow x^i$, $p_j \rightarrow -i\hbar\frac{\partial}{\partial x^j}$ works fine. But what happen if Q is not linear ?
\item How do we quantize observables which involve higher powers of $q^i, p_j$, as for example $f(q^i,p_j) = (q^i)^n(p_j)^m$ when $n+m \ge 3$?
\item How should we quantize more general phase spaces, which are the symplectic manifolds  not necessarily cotangent bundles?
\end{enumerate}
 
As we are currently interested in new developments in quantization methods \cite{lg1, lg2, lg3}, the goal of this paper is to highlight through simple models the benefits of affine quantization.

In section (\ref{sec2}), we revisit a method of quantization by J. R. Klauder, the affine quantization, then in section (\ref{sec3}) we formulate and solve the free particle and the harmonic oscillator. Concluding remarks are given in section (\ref{sec4}). 

\section{Affine Quantization}\label{sec2}

While in confinement due to COVID-19, our attention has been drawn on a recent published paper of J. R.  Klauder on {\it The benefits of Affine Quantization} \cite{jklau1}. Our motivation is due to the fact that there is a difficulty with canonical quantization when it comes to configuration spaces other then $\mathbb{R}^n$. Consider, for example, a particle that is restricted to move on the positive real line. The configuration space is $\mathbb{Q} = \mathbb{R}^+$. It seems reasonable to use the position q and momentum p as classical observables, which satisfy the usual commutations relations. However, when we try to represent these by operators $\hat q \equiv q$ and $\hat p \equiv -i\hbar\frac{\partial }{\partial q}$, it turns out that the momentum operator $\hat p$ is not self-adjoint on the Hilbert space $\mathcal{H} = L^2(\mathbb{R}^+, dq)$. Thus a straightforward application of Dirac's canonical quantization recipe is impossible. 

Our goal is to apply the method of affine quantization to study 
the free particle and the harmonic oscillator that would serve as a test in order to study a toy  model of a massive Klein Gordon field coupled to an harmonic oscillator at the the boundary considering the half line.
In this section, we revisit the affine quantization method from some previous works of J. R. Klauder \cite{jklau2,jklau3,jklau4}.

Let us start with a single degree of freedom, the classical phase space variables $p$ and $q$ are real satisfying a standard Poisson bracket 
\begin{equation}\label{eq21}
\{q, p\} = 1 ,
\end{equation}
 multiplying by $q$ the equation (\ref{eq21}) we get 
 \begin{equation}
 q\{q,\:p\} = q, 
 \end{equation}
that is equivalent to $\{q, \: pq\} = q$, setting $d = pq$, we have 
\begin{equation}
\{q,\: d\} = q.
\end{equation}
The two variables $d$ and $q$ form a Lie-algebra and are worthy of consideration as new pair of classical variables even though they are not canonical coordinates. It is also possible to restrict $q$ to $q> 0$ or $q<0$ consistent with $d$. The variable $d$ acts to dilate $q$ and not to translate $q$ as the variable $p$ does. 

For the case of the single degree of freedom above, the canonical quantization involves $\hat q$ and $\hat p$ which are self adjoint operators that satisfy the canonical commutation relation
\begin{equation}
[\hat q,\:\hat p] = i\hbar I_d.
\end{equation}
From the canonical quantization, it follows that 
\begin{equation}\label{eq22}
\hat q [\hat q,\:\hat p] = [\hat q,\:\hat q\hat p] = 
[\hat q, \frac{(\hat q\hat p + \hat p\hat q)}{2}]\equiv 
[\hat q, \hat d] = i\hbar \hat q , 
\end{equation}
where the dilation operator is define as $\hat d\equiv 
(\hat p \hat q + \hat q\hat p)/2 $ is self adjoint.
The ope\-ra\-tor $\hat d$ is called the dilation operator because it 
dilates $\hat q$ rather than translates $\hat q$ as $\hat p$ does, in particular 
\begin{equation}
e^{\frac{iq\hat p}{\hbar}}\hat q e^{-\frac{iq\hat p}{\hbar}} 
= \hat q + q I_d ,
\end{equation}
while
\begin{equation}\label{dilop}
e^{i\ln(|q|)\hat d/\hbar} \hat q e^{-i\ln(|q|)\hat d/\hbar}
= |q| \hat q = q |\hat q|.
\end{equation}
In the second relation in equation (\ref{dilop}), $ q\neq 0$, and $q$ as well as $\hat q$ are normally chosen to be dimensionless. According to the equation 
(\ref{eq22}), the existence of canonical operators guarantees the existence of affine operators.
If $\hat q > 0$ or ($\hat q < 0$), then the operator $\hat p$ cannot be made self-adjoint, however in that case, both of the operators $\hat q$ and $\hat d$ are self adjoint. As usual 
$\hat q$ and $\hat p$ are irreducible, but $\hat q$ and $\hat d$ are reducible. There are three inequivalent irreducible representations; one with 
$\hat q > 0$, one with $\hat q < 0$ and one with $\hat q = 0$ and all three involve representations that are self-adjoint. The first two irreducible choices are the most interesting and, for the present, we focus on the choice $\hat q > 0$. 
 
\section{Testing some models}\label{sec3}

\subsection{The free particle}

The simplest Hamiltonian one can envisage is the free particle on the half line. The Hamiltonian reads 
\begin{equation}
\mathcal{H}_f(x, p_x) = \frac{1}{2m}p_x^2,
\end{equation}
with $\{x, p_x\} = 1, \quad x > 0$, 
in terms of the affine variables as described in section (\ref{sec2}) we may rewrite
\begin{equation}
\mathcal{H}_f(x, d_x) = \frac{1}{2m}d_x x^{-2}d_x,
\end{equation}
where the variable $d_x$ is the dilation variable $d_x = p_x x$
and $\{x, d_x\} = x$.
By mean of canonical quantization where the affine variables $x,\; d_x$ are respectively promoted to operators $\hat x, \hat d_x$ the 
corresponding Hamiltonian for the free particle reads
\begin{equation}
\hat \mathcal{H}_f (\hat x,\hat d_x) = \frac{1}{2m}\hat d_x (\hat x)^{-2} \hat d_x,
\end{equation}
where $\hat d_x$ stands for the dilation operator and $[\hat x, \; \hat d_x] = i\hbar \hat x$. We have the representation
$\hat d \equiv -i\hbar ( x\partial_x + 1/2)$ and $\hat x\equiv x$, with $x > 0 $. The time independent eigenvalue equation can be written as 
\begin{equation}\label{pb1}
( \frac{1}{2m}\hat d_x (\hat x)^{-2}\hat d_x )\phi(x) = E\phi(x),
\end{equation}
that is equivalent to 
\begin{equation}\label{pb2}
\left( -\frac{\hbar^2}{2m}\frac{d^2}{dx^2}
 +\frac{\hbar^2}{2m}\frac{3}{4}x^{-2} \right)\phi(x) = E\phi(x).
\end{equation}
We are then interested in solving the problem (\ref{pb2}).
If we divide by $-\frac{\hbar^2}{2m}$ and setting $k^2 = \frac{2m E_k}{\hbar^2}$, where we assume $E_k > 0$ and label $\alpha = 3/4$, the equation (\ref{pb2}) takes the form
\begin{equation}\label{pb3}
\phi''_k(x) = \left( \frac{\alpha}{ x^{2}} - k^2\right) \phi_k(x).
\end{equation}
Let's consider the change of variable variable $x = k^{-1} y$. The equation in (\ref{pb3}) is rewritten in terms of the new variable $y$ as follows 
\begin{equation}
\phi''(y) = \left(\frac{\alpha}{y^2}  -1 \right)\phi(y).
\end{equation}
Setting $\phi(y) = y^{1/2}\varphi(y)$ and reminding that $\alpha = 3/4$, we obtain the ordinary differential equation 
\begin{equation}
\varphi''(y) + \frac{1}{y}\varphi'(y) + \left(1 -\frac{1}{y^2}\right)\varphi(y) = 0,
\end{equation}
that is a variant of the Bessel's equation and the solution is defined by the Bessel function of order one, $J_1(x)$.
A continuum of eigenfunctions exist for the problem (\ref{pb3}), hence 
for problem (\ref{pb2}) as
\begin{equation}
\phi_k(x) = (kx)^{\frac{1}{2}}J_1(kx), \:\:
E_k = \frac{k^2\hbar^2}{2m}, \: k >0\;.
\end{equation}
that are satisfying the important closure relation 
\begin{equation}
\int_0^\infty\phi_k(x)\phi_k(y) dk = \delta(x-y),\: k >0\;.
\end{equation}
since the order of the Bessel function is greater than $(-1/2)$. The details about that property can be found in reference \cite{weber}.

\subsection{The harmonic oscillator}

We consider the one dimensional harmonic oscillator represented by the classical Hamiltonian 
\begin{equation}\label{311}
H_o(x,p_x) = \frac{1}{2m}p_x^2 + \frac{1}{2}m \omega^2 x^2,
\end{equation}
where $(p_x,x)\in \mathbb{R}\times \mathbb{R}^+$, that means $x >0$, with $\{x, p_x\} = 1$. Our aim is to test the affine quantization that has already been partially considered in \cite{jklau4}. Let us first determine the classical affine variables as in the section (\ref{sec2}). We set $ d_x = p_xx$, also called the dilation variable and the new variables called affine variables are $x$ and $d_x$ that satisfy the relation 
$\{x, d_x\} = x$. The Hamiltonian in equation (\ref{311}) can be then rewritten in terms of the affine variables as 
\begin{equation}
H_a(x,d_x) = \frac{1}{2m}d_x(x^{-2})d_x + \frac{1}{2}m\omega^2 x^2.
\end{equation}
By mean of the affine quantization, the classical affine variables are promoted as operators 
\begin{equation}
d_x \rightarrow \hat d_x; \quad x \rightarrow \hat x,
\end{equation}
and the corresponding quantized Hamiltonian is 
\begin{equation}
\hat H_a (\hat x, \hat d_x)= \frac{1}{2m}\hat d_x (\hat x)^{-2} \hat d_x + \frac{1}{2}m\omega^2 \hat x^2.
\end{equation}
The affine operators satisfies the commutation relations
\begin{equation}
[\hat x, \hat d_x\;] = i\hbar \hat x,
\end{equation}
and act as follows
\begin{equation}
\hat x\psi (x,t) = x\psi(x,t); \quad x> 0,\; \hat x >0,
\end{equation}
and 
\begin{equation}
\hat d_x \psi (x,t) = -i\hbar \left(x\frac{\partial}{\partial x}+\frac{1}{2}\right)\psi(x,t),
\end{equation}
where the wave functions are normalized 
\begin{equation}
\int_0^\infty \vert \psi(x,t)\vert^2dx = 1.
\end{equation}
We can then solve for the Schr\"odinger equation
\begin{equation}\label{312}
i\hbar\frac{\partial\psi(x,t)}{\partial t} = \hat H_o\psi(x,t).
\end{equation}
Since we are in presence of autonomous system we may set the Ansatz
\begin{equation}
\psi(x,t) = e^{-itE/\hbar}\phi(x),
\end{equation}
and then the corresponding time-independant eigenvalue equation is 
\begin{equation}
\hat H_a(\hat x, \hat d_x) \phi(x) = E \phi(x),
\end{equation}
that is  explicitly the equation
\begin{equation}\label{eq23}
\left( -\frac{\hbar^2}{2m}\frac{d^2}{dx^2} + 
\frac{3\hbar^2}{8m}\frac{1}{x^2} + \frac{1}{2}m\omega^2 x^2 \right)\phi(x) = E\phi(x),
\end{equation}
and we can rewrite as 
\begin{equation}\label{eq24}
\left[ -\frac{d^2}{dx^2} + \frac{3}{4}\frac{1}{x^2} + \frac{m^2\omega^2}{\hbar^2} x^2 \right] \phi(x) = \frac{2mE}{\hbar^2}\phi(x).
\end{equation}
For matter of simplication, let's set the following parameters
\begin{equation}
\lambda = \frac{m\omega}{\hbar}; \quad k^2 = \frac{2m E}{\hbar^2};\quad \alpha = \frac{3}{4},
\end{equation}
and the equation (\ref{eq24}) becomes 
\begin{equation}
\left[ -\frac{d^2}{dx^2} + \lambda^2 x^2 + \frac{\alpha}{x^2} \right]\phi(x) = k^2 \phi(x).
\end{equation}
A standard asymptotic analysis for $x\rightarrow \infty,\: x\rightarrow 0$, requires the following Ansatz for the wave function
\begin{equation}\label{ans}
\phi(x) = x^{\beta +1} e^{-\frac{\lambda}{2} x^2}v(x),
\end{equation}
where  the constant $\beta $ and $\alpha$ are related  by $\alpha = \beta (\beta +1)$ and $v(x)$ is unknown function. The relation between $\alpha$ and $\beta$ gives two possible values of $\beta$ that are $ \beta_+ = +1/2$ and $\beta_- = -3/2$. We choose not to specify for the moment the values of  $\beta$. From the Ansatz in equation (\ref{ans}), an equation for the unknown function $v(x)$ is given by 
\begin{equation}\label{eq25}
v''(x) +\left[ 2(\beta +1) x^{-1} -2\lambda x\right] v'(x) + 
\left[ k^2 -\lambda(2\beta +3)\right] v(x) = 0.
\end{equation}
Making the change of variable $y = \lambda x^2$ in equation (\ref{eq25}), we obtain the differential equation
\begin{equation}\label{eq26}
y v''(y) + \left[ (\beta +3/2)-y \right]v'(y) + 
\left[\frac{k^2}{4\lambda} -\frac{1}{4}(2\beta + 3)
\right]v(y) = 0 .
\end{equation}
The general solution of the equation (\ref{eq26}) also known as 
Kummer's differential equation, can be expressed in terms of confluent hypergeometric functions
\begin{eqnarray}\nonumber
v(y) &=& A\; \mbox{}_1F_1\left(\frac{1}{2}(\beta +3/2)-\frac{1}{2}\mu,\; \beta + 3/2,\; y \right)\\\label{eq27}
&+& B\; y^{-(\beta +1/2)}\mbox{}_1F_1\left(\frac{1}{2}(-\beta +1/2)-\frac{1}{2}\mu,\; \frac{1}{2}-\beta,\; y\right),
\end{eqnarray}
where $\mu  = k^2/(2\lambda)$, and $\mbox{}_1F_1(a,c,y)$ denotes the confluent hypergeometric function which has the following series representation
\begin{equation}
\mbox{}_1F_1 (a,c,y) = 1 +\frac{a}{c} y + \frac{a(a+1)}{c(c+1)}\frac{y^2}{2!} + \ldots
\end{equation}
Due to the asymptotic behavior of the confluent hypergeometric function given by 
\begin{equation}
\mbox{}_1F_1 (a,c,y) \sim e^yy^{a-c},
\end{equation}
which implies divergencies of both of the terms in equations (\ref{eq27}) and then the impossibility to normalize the wave function,  we impose the following conditions 
$\frac{1}{2}(\beta + 3/2) - \frac{1}{2}\mu = -n$ or 
$\frac{1}{2}(-\beta + 1/2)-\frac{1}{2}\mu = -n, \: n= 1,2,\ldots$

For the first condition, that is 
$\frac{1}{2}(\beta + 3/2) - \frac{1}{2}\mu = -n$, the eigenfunctions of the equation 
(\ref{eq24}) have the form 
\begin{equation}
\phi_n(x) = A_n x^{\beta+1}e^{-\frac{m\omega}{2\hbar}x^2}
\mbox{}_1F_1\left(-n, \beta + 3/2, \frac{m\omega}{\hbar}x^2 \right),
\end{equation}
with the energy levels of the form
 $E_n = \hbar\omega(2n + \beta +3/2 ),\: n=1,2,\ldots $, 
where the constants $A_n$ have to be determined using the normalization conditions. We have for $\beta = + 1/2$, 
the solution
 \begin{equation}\label{hyp1}
\phi_n(x) = A_n x^{3/2}e^{-\frac{m\omega}{2\hbar}x^2}
\mbox{}_1F_1\left(-n, 2, \frac{m\omega}{\hbar}x^2 \right),
 E_n = 2(n+1)\hbar\omega, \: n= 0, 1,2,\ldots
\end{equation}
The value of $\beta = -3/2$ is skipped as it leads to undefined 
confluent hypergeometric function and $n= 0$ can be included in the equation (\ref{hyp1}).

For the second condition, that is 
 $\frac{1}{2}(-\beta + 1/2)-\frac{1}{2}\mu = -n$, the eigenfunctions of the equation (\ref{eq24}) have the form 
 \begin{equation}
 \phi_n(x) = B_n (\frac{m\omega}{\hbar})^{-(\beta+1/2)} x^{-\beta}
 e^{-\frac{m\omega}{2\hbar}x^2}\mbox{}_1F_1\left(-n, \frac{1}{2}-\beta, \frac{m\omega}{\hbar}x^2\right),
 \end{equation}
with $E_n = (2n -\beta +\frac{1}{2})\hbar\omega,\: n= 1,2,\ldots $, 
where the constants $B_n$ have to be determined using the normalization conditions. For  $\beta = -3/2$ we have 
\begin{equation}\label{hyp4}
 \phi_n(x) =  (\frac{m\omega}{\hbar}) B_n x^{3/2}
 e^{-\frac{m\omega}{2\hbar}x^2}
 \mbox{}_1F_1\left(-n, 2, \frac{m\omega}{\hbar}x^2\right), E_n = 2(n+1)\hbar\omega,\:n = 0,1,2,\ldots
 \end{equation}
 The value of $\beta = +1/2 $ leads in this case to an 
 undefined confluent hypergeometric function and $n= 0$ can be included in equation (\ref{hyp4})
 
The constant $A_n$ and $B_n$ in equation  (\ref{hyp1}) and respectively in equation (\ref{hyp4}) are determined from the 
normalization condition 
\begin{equation}\label{normeq}
\int_0^\infty \phi_n^2(x) dx = 1
\end{equation}
In order to compute the integral (\ref{normeq}) for each case,
we use a lemma which is a generalization of formula $f6$ in the book {\it Quantum Mechanics } by Landau and Lifshitz \cite{landau}.

{\it Lemma :
For $\gamma > 0$, and $m, n = 0,1,2,\ldots$
\begin{equation}
\int_0^\infty x^{2\gamma -1}e^{-\beta x^2}
\mbox{}_1F_1(-n;\gamma,\beta x^2 )
\mbox{}_1F_1(-m;\gamma,\beta x^2 )dx 
= \frac{1}{2} n! \frac{\Gamma(\gamma)}{\beta^\gamma(\gamma)_n}\;
\delta_{mn},
\end{equation}
where $\delta_{mn} = 0$ for $m\neq n$ and $1$ for $m = n$, and 
$(\gamma)_n$ is the Pochhammer symbol $(\gamma)_n \equiv \Gamma( \gamma + n)/\Gamma (\gamma)$}.

The constants $A_n$ and $B_n$ respectively in the equations (\ref{hyp1}) and (\ref{hyp4}) are explicitly given by 
\begin{equation}
A_n = (\frac{m\omega}{\hbar})\sqrt{\frac{2\;(2)_n}{n!\;\Gamma(2)}}\;, \quad B_n = \sqrt{\frac{2\;(2)_n}{n!\;\Gamma(2)}}, \: 
n= 0, 1,2, \ldots,
\end{equation}
 which are simplified using $(2)_n \equiv \Gamma (2 +n)/\Gamma (2)$ to
 \begin{equation}
 A_n = (\frac{m\omega}{\hbar})\sqrt{2(n+1)}\;,\quad B_n = \sqrt{2(n+1)}\;.
 \end{equation}

To summarize the eigenfunctions and the eigenvalues of the equation (\ref{eq24}) are respectively given by
\begin{equation}\label{hypf1}
\phi_n(x) = \sqrt{2(n+1)} (\frac{m\omega}{\hbar})x^{3/2}e^{-\frac{m\omega}{2\hbar}x^2}
\mbox{}_1F_1\left(-n, 2, \frac{m\omega}{\hbar}x^2 \right),\:
n= 0,1,2\ldots
\end{equation}
\begin{equation}\label{hypen1}
E_n = 2(n+1)\hbar\omega,\; n= 0, 1, 2, \ldots
\end{equation}
The eigenfunctions as we can see in the equation (\ref{hypf1}) satisfy the orthogonality conditions and are different from the ones of the full harmonic oscillator. They are similar to the ones of the family of harmonic oscillators to which a singular repulsion at the origin is added. The energy eigenvalues in the equation (\ref{hypen1}) are countably and equally spaced.
 
\section{Concluding remarks}\label{sec4}

We have revisited the procedure of affine quantization introduced by J. R. Klauder. Our motivation is to better understand this quantization method on nontrivial phase spaces where canonical quantization fails. We have tested the procedure for the simple case of the free particle and the case of the harmonic oscillator, both on the half-line. 

For the case of the free particle in the upper half-line, the equivalent of the Hamiltonian in the affine coordinates turns out to be the case of a particle in a square inverse potential $V(x)= \alpha/x^2$. This kind of problem has been considered in the literature \cite{steve, griff, coon} and the case of $-1/4 < \alpha < 0$ has been explicitly discussed in \cite{steve}. There is a richness in the inverse square potential in quantum mechanics due to the connections to diverse physical phenomena \cite{kam,krae, kap}.

For the case of the harmonic oscillator on the half-line, usually called the half-harmonic oscillator, the equivalent of the Hamiltonian in terms of the affine coordinates is the case of model with the potential $V(x) = (1/2) m\omega^2x^2 + \alpha/x^2, \; x >0 $, that is a well-known model in the literature.  In fact, the family of quantum Hamiltonians known as spiked harmonic oscillators is given by the general Hamiltonian operator $ H= -d^2/dx^2 + x^2 + \alpha/x^2$ acting on the Hilbert space $L^2 (0, \infty)$. The name of the operator derives from the graphical shape of the full potential $ V(x)=  x^2 + \alpha/x^2 $ which shows a pronounced peak near the origin for $\alpha > 0$.
The spiked harmonic oscillator has drawn the attention of many authors since the publication of the pioneering paper of J. R. Klauder \cite{jrklau}, four decades ago. It represents the simplest model of certain realistic interaction potentials in atomic, molecular and nuclear physics, and second and also due to its interesting intrinsic properties from the viewpoint of mathematical physics \cite{spho1,spho2}.
The method used in section (\ref{sec3}) is similar to the one discussed in the problem of quantization of the systems with a position-dependent mass \cite{mikha3}.

We have been curious about the possibility of getting back to canonical quantization from affine quantization. Is - it possible to recover the solutions of the usual harmonic oscillator on the line from the solutions of the half harmonic oscillator ?
In order to analyse that question, we consider a positive real $b\ge 0$, where $-b < x$. In the situation of the half-line, the 
endpoint and the symmetry point coincide to $0$. Now we imagine the situation in which we save the symmetry point that remains $0$ while the endpoint can move toward negative infinity. The situation is described by the Hamiltonian 
\begin{equation}
\hat H_a(\hat x, \hat d_x) = \frac{1}{2m}\left(
(\hat d_x +b\hat p_x)(\hat x +b)^{-2}(\hat d_x +b\hat p_x)\right)
+ \frac{1}{2}m\omega\hat x^2, 
\end{equation}
 where $\hat d_x$ is the dilation operator defined in the previous sections and 
 \begin{equation}
 [\hat x, \hat p_x] = i\hbar, \quad
 [\hat x, \hat d_x] = i\hbar \hat x\;.
 \end{equation}
The corresponding time independent Schr\"odinger equation is given by 
\begin{equation}\label{hext}
\left[ -\frac{d^2}{dx^2} + \frac{3}{4}\frac{1}{(x+b)^2} + \frac{m^2\omega^2}{\hbar^2} x^2 \right] \phi(x) = \frac{2mE}{\hbar^2}\phi(x),
\end{equation} 
When $b = 0$, we recover the problem in section (\ref{sec3}), and 
when $b\rightarrow +\infty$ the equation (\ref{hext}) tends to the one of the harmonic oscillator on the line.
While the equation (\ref{hext}) is not obvious to solve analytically, we guess that the limit  $b\rightarrow +\infty$ leads from affine to canonical and should lead to usual even and odd about the symmetry point $x=0$ eigenfunctions.
The study of this problem is presently under investigation as another example of how affine quantization deals with harmonic oscillator problems that canonical quantization cannot resolve \cite{lrgoub}.

We are also interested in considering more toy models for instance the case of the  model that consists of a massive Klein Gordon field in $1+1$ dimensions restricted to the left half-line by a boundary  and coupled to a harmonic oscillator at that boundary, thus introducing some additional degrees of freedom. The classical Hamiltonian is defined by 
\begin{eqnarray}\nonumber
H &=& \int_{-\infty}^0 dx\left(
\frac{1}{2}\pi(x,t)^2 +\frac{1}{2}\left(\partial_x\phi(x,t) \right)^2 +\frac{1}{2}\mu^2\phi(x,t)^2
\right) \\
&+& \beta \phi(0,t)q(t) +\frac{1}{2} m\omega^2q(t)^2
+\frac{1}{2m}p(t)^2.
\end{eqnarray}
We hope to report on that problem soon \cite{laure}.

As pointed out in reference \cite{jklau4}, gravity does not fit well with canonical quantization and affine quantization may be an alternative procedure. So we are also interested in understanding how the affine quantization can help in quantum gravity.

\vspace{0.3cm}
\noindent {\bf Acknowledgements}: 

L. Gouba would like to gratefully thank Professor John R. Klauder for the comments, the suggestions and the ongoing discussion on the paper and its extensions.

L. Gouba would like to gratefully thank Professor Mikhail S. Plyushchay for useful and instructive discussions.

L. Gouba would like to thank the Family and the Friends for their presence and their constant support especially during the COVID-19 lockdown.


\vspace{0.3cm}

\begin{thebibliography}{50}

\bibitem{mikha1}{Gamboa J and Plyushchay S M, 1998 {\it Classical anomalies for spinning particles}, Nucl. Phys. B,
vol 512 pp 485-504}

\bibitem{mikha2}{ Correa F , Jakubsk\'y  V and Plyushchay S M, 2009 {\it Aharonov-Bohm effect on AdS2 and nonlinear supersymmetry of reflectionless P\"oschl-Teller system },  Annals Phys. vol 324 pp 1078-1094}

\bibitem{pamd}{Dirac P A M, 1925 {\it Proc. of the Royal Society of London A : Mathematical, Physical and Engineering Sciences } vol 109 pp 642}

\bibitem{dir}{ Dirac P A M, 1930 {\it The principles of Quantum Mechanics} (Oxford: Oxford University Press) pp 157-162,281-285}

\bibitem{wey}{Weyl H, 1950 {\it The Theory of Groups and Quantum Mechanics} (New York: Dover Publishing Compagny, Inc.) pp 274-275}

\bibitem{vneu}{ von Neumann J, 1955 {\it Mathematical Foundations of Quantum Mechanics} (Princeton: Princeton University Press) }

\bibitem{born}{ Born M and Jordan P, 1925 On quantum mechanics {\it Zs f. Phys.} {\bf 34} 858}

\bibitem{agar}{Agarwal B S and Wolf E, 1970 Calculus for functions of noncommutating operators and general phase space methods in quantum mechanics {\it Phys. Rev. D} {\bf{2}} 2206-25}

\bibitem{geo1}{ Weyl H K H, 1927 Quantenmechanik und gruppentheorie
 {\it Z. Phys. } {\bf{46}} 1- 46 }

\bibitem{geo2}{Groenewold H J, 1946 {\it  Physica} 
{\bf{12}} 405-60 }

\bibitem{geo3}{Souriau J M, 1966 Quantification g\'eometrique  {\it Commun.  math.  Phys.}  {\bf{1}} 374}

\bibitem{geo4}{Kostant B, 1970 {\it Lecture Notes in Mathematics}  (Berlin: Springer-Verlag) Vol 170 pp 87-208}

\bibitem{fey}{Feynman R P, 1942 {\it Principles of Least Action in Quantum Mechanics} (Thesis Feynman)}

\bibitem{def2}{ Bayen F, Flato M, Fronsdal C, Lichnerowicz A and Sternheimer D, 1978 Deformation theory and quantization I, II
{\it  Ann. Phys. (NY) }{\bf 111} 61-110, 111-151}

\bibitem{def3}{Kontsevich M, 1997 Deformation quantization of Poisson manifolds {\it Preprint q-alg/9709040 }}

\bibitem{def4}{ Bordemann M and Waldmann S, 1998 Formal GNS construction and states in deformation quantization {\it Comm. Math. Phys.} {\bf 195} 549-583}

\bibitem{deff}{ Bordemann M, Neumaier N and Waldmann S,
1998 Homogeneous Fedosov star products on cotangent bundles I: Weyland standard ordering with differential operator representation {\it Comm. Math. Phys.} {\bf 198} 363-396}

\bibitem{def5}{Dito G and Sternheimer D, 2002 Deformation quantization: genesis, developments and metamorphoses
{\it Preprint math.QA/0201168}}

\bibitem{kbt1}{ Klauder J R, 1967 Weak correspondence principle  {\it J. Math. Phys.} {\bf 8} 2392}

\bibitem{kbt2}{Berezin F A, 1974 Quantization {\it Izv. Akad. Nauk SSSR Ser. Mat.} {\bf 38} 1116-1175}

\bibitem{ber}{Berezin F A, 1975 General concept of quantization 
{\it Comm. Math. Phys.} {\bf 40} 153-174} 

\bibitem{kbt3}{Coburn L A, 1994 {\it Algebraic Methods in Operator Theory} ( Boston: Birkh\"auser Springer) pp 101-108}

\bibitem{lg1}{Gouba L, 2019 {\it New Trend in Quantization Method}, arXiv:1912.05388 [math-ph]}

\bibitem{lg2}{Gouba L, 2019 Beyond Coherent States Quantization, {\it J. Phys.: Conf. Ser.1416 012012}, Open access }

\bibitem{lg3}{Gouba L, 2018 Dirac's Method for the Two-Dimensional Damped Harmonic Oscillator in the Extended Phase Space, {\it Mathematics 2018, 6, 180} (This article belongs to the Special Issue Time and Time Dependence in Quantum Mechanics), Open Access}

\bibitem{jklau1}{Klauder J R,  2020 The benefits of affine quantization, {\it Journal of High Energy Physics, Gravitation and Cosmology}, 175-185}

\bibitem{jklau2}{Klauder J R, 2012 Enhanced quantization: a primer, {\it J. Phys. A: Math. Theor.} {\bf 45} 285304 }

\bibitem{jklau3}{Klauder J R, 2016 When canonical quantization fails, how to fix it, arXiv:1611.02107 [quant-ph] }

\bibitem{jklau4}{Klauder J R 2020, Quantum gravity made easy, 
Journal of High Energy Physics, Gravitation and Cosmology, {\bf 6}, 90-102}

\bibitem{weber}{Arfken G B and  Werber H J, 2005 {\it  Mathematical Methods for Physics}, 6eme Ed(Amsterdam, Elsevier) pp 696}


\bibitem{landau}{Landau L D and Lifshitz E M, Quantum Mechanics: non-relativistic theory (Pergamon, London, 1981)}

\bibitem{steve}{ Paik S T, 2018 Teaching renormalization, scaling, and universality with an example from quantum mechanics, {\it J Phys. Commun 2015016}}


\bibitem{griff}{Essin A M, Griffiths D J, 2006 Quantum mechanics of the $1/x^2$ potential {\it Am. J. Phys. 74 109.17}}

\bibitem{coon}{Coon S A and Holstein B R, 2002 Anomalies in quantum mechanics: the $1/r^2$ potential, Am. J. Phys. 70513 }

\bibitem{kam}{Camblong H E, Epele L N, Fanchiotti H and Garcia Canal C A, 2001 Quantum anomaly in molecular physics Phys. Rev. Lett. 87 220402}

\bibitem{krae}{ Kraemer T et al, 2006 Evidence for Efimov quantum states in an ultracold gas of cesium atoms {\it Nature 
440 315–8} }

\bibitem{kap}{Kaplan D B, Lee J-W, Son D T and Stephanov M A, 2009 Conformality lost {\it Phys. Rev. D80 125005}}

\bibitem{jrklau}{Klauder J R, 1978 Continuous and discontinuous perturbations, Science 199 735}


\bibitem{spho1}{Aguilera - Navarro V C, Guardiola R, 1991 Nonsingular spiked harmonic oscillators, {\it Journal of Mathematical Physics, 32, 2135} }

\bibitem{spho2}{Hall R L, Saad N and von Keviczky A B, 2001 Spiked Harmonic Oscillator, {\it Journal of Mathematical Physics, 43 (1)}}

\bibitem{mikha3}{Bravo R and Plyushchay S M, 2016 {\it Position-dependent mass, finite-gap systems} Phys. Rev. D, vol 93, number 10, pp 105023}

\bibitem{lrgoub}{Gouba L, The b-problem for the half harmonic oscillator, {\it in preparation}}

\bibitem{laure}{ Gouba L,  Affine quantization of the massive Klein Gordon field coupled to an oscillator harmonic at the boundary, {\it in preparation}}

\end{thebibliography}
\end{document}